# Digitally-assisted photonic analog domain self-interference cancellation for in-band full-duplex MIMO systems via LS algorithm with adaptive order

Moxuan Han[a,b], and Yang Chen[a,b,*]

[a] Shanghai Key Laboratory of Multidimensional Information Processing, East China Normal University, Shanghai, 200241, China
[b] Engineering Center of SHMEC for Space Information and GNSS, East China Normal University, Shanghai, 200241, China
[*] ychen@ce.ecnu.edu.cn

**ABSTRACT**
A digitally-assisted photonic analog domain self-interference cancellation (SIC) and frequency downconversion method is proposed for in-band full-duplex multiple-input multiple-output (MIMO) systems using the least square (LS) algorithm with adaptive order. The SIC and frequency downconversion are achieved in the optical domain via a dual-parallel Mach–Zehnder modulator (DP-MZM), while the downconverted signal is processed by the LS algorithm with adaptive order that is used to track the response of the multipath self-interference (SI) channel and reconstruct the reference signal for SIC. The proposed method can overcome the reconstruction difficulty of the multipath analog reference signal for SIC with high complexity in the MIMO scenario and can also solve the problem that the order of the reference reconstruction algorithm is not optimized when the wireless environment changes. An experiment is carried out to verify the concept. 30.2, 26.9, 23.5, 19.5, and 15.8 dB SIC depths are achieved when the SI signal has a carrier frequency of 10 GHz and baud rates of 0.1, 0.25, 0.5, 1, and 2 Gbaud, respectively. The convergence of the LS algorithm with adaptive order is also verified for different MIMO multipath SI signals.

## 1. Introduction

In-band full-duplex (IBFD) communication [1,2] theoretically doubles the efficiency of spectrum usage over frequency-division duplex and time-division duplex communication by simultaneously transmitting and receiving signals in the same frequency band. However, it suffers from strong self-interference (SI) caused by the leakage from the transmitting antenna, which seriously affects the reception of the signal of interest (SOI).

The SI signal can be eliminated using mature electrical-based methods [3–5]. These technologies, however, are frequently limited in operating frequency and bandwidth due to the electronic bottleneck. Because microwave photonic technology [6] can overcome the inherent defects of the electrical-based solutions, photonics-assisted self-interference cancellation (SIC) has been widely studied in the past few years [7–14]. In [7–11], the strong direct-path SI signal was taken into consideration, and the SIC methods proposed in [7,9,11] can also overcome the influence of fiber dispersion when the SIC system was combined with fiber transmission. In [12–14], different methods for multipath SIC were proposed, whose essence was to replicate the single-path SIC schemes in different dimensions to achieve independent cancellation of SI signals on different paths. The complexity of the system is closely related to the number of paths of the multipath SI signal, which limits the feasibility of such methods in practical applications. Therefore, it is necessary to study photonics-assisted methods for multipath SI signal with a simple structure and the ability to estimate the response of the multipath SI channel.

In [15], the least square (LS) algorithm was introduced into a photonics-assisted SIC system based on two electro-absorption modulated lasers to estimate the response of the time-varying multipath SI channel and then pre-compensate the reference signal in the frequency domain to suppress the multipath SI signal in the optical domain. This method achieved an SIC depth of 40 dB at 900 MHz bandwidth. In [16], an adaptive photonic analog multipath SIC method with a 32-dB SIC depth at 2.7 GHz bandwidth was proposed, which employed the least mean square algorithm-based adaptive filter as a pre-equalizer and a dual-drive Mach–Zehnder modulator (DD-MZM) for suppressing the multipath SI signal in the optical domain. In [17], we proposed a multipath SIC and frequency downconversion method based on a DD-MZM and the recursive least square (RLS) adaptive algorithm. In this scheme, one reference signal was directly tapped from the transmitter to cancel the strong direct-path SI signal, while the other reference signal was reconstructed by the RLS algorithm to cancel the residual direct-path SI signal and multipath SI signal. Nevertheless, although the above methods realize the multipath analog SIC in much simpler structures with the assistance of digital algorithms, these methods cannot adjust the order of the algorithm adaptively with the variation of the multipath SI channel. When the multipath SI delay spread varies greatly, it is difficult to balance the performance and complexity of the algorithm.

Furthermore, multiple-input multiple-output (MIMO) technology can also improve the capacity and spectrum utilization of wireless communication systems without increasing its bandwidth. When combined with IBFD technology, each receiving antenna in the MIMO receiver will receive multiple different SI signals of the same frequency. The SI problem in this scenario is much more complicated and has not been thoroughly studied using the photonics-based method.

In this letter, a digitally-assisted photonic analog domain SIC and frequency downconversion method is proposed for IBFD MIMO systems using the LS algorithm with adaptive order. Unlike the algorithms with fixed order in [15–17], the LS algorithm adaptively adjusts the order to better estimate the response of the multipath SI channel and track the variation of the multipath SI channel, so as to achieve a balance in the complexity and performance of the algorithm for reconstructing the reference signal. Moreover, photonics-assisted SIC is first demonstrated for IBFD MIMO systems, in which different transmitting antennas transmit different data streams for increasing its capacity.

## 2. Principle

The schematic of the proposed system is shown in Fig. 1, which mainly consists of the MIMO transmitter, the photonic analog domain SI canceler and frequency downconverter, and the reference constructor based on digital signal processing (DSP). For each MIMO receiving antenna, the SIC structure is the same. Thus, in the experiment, only two transmitting antennas and one receiving antenna are used to demonstrate the concept in this work.

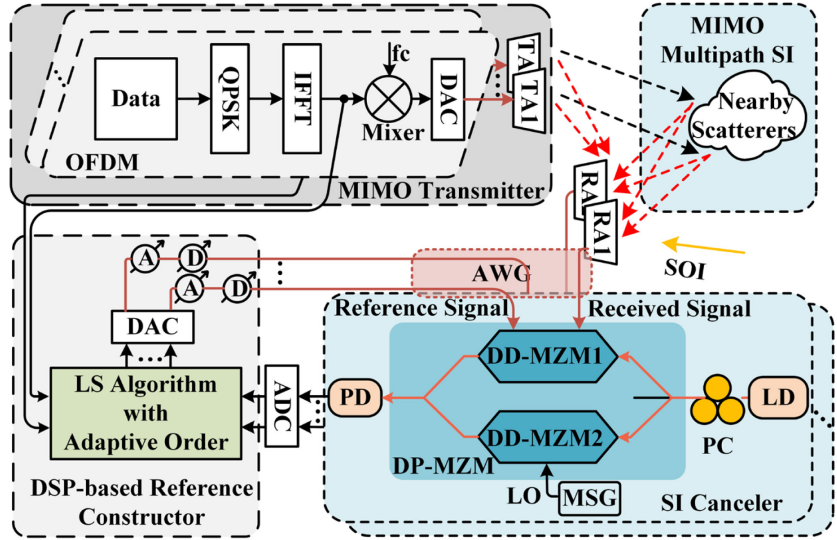

Fig. 1. Schematic of the proposed digitally-assisted photonic analog domain SIC and frequency downconversion system. TA, transmitting antenna; RA, receiving antenna; DD-MZM, dual-drive Mach–Zehnder modulator; DP-MZM, dual-parallel Mach–Zehnder modulator; PC, polarization controller; MSG, microwave signal generator; AWG, arbitrary waveform generator; LO, local oscillator; LD, laser diode; PD, photodetector; A, tunable electrical attenuator; D, tunable electrical delay line.

A 15.5-dBm continuous-wave light wave from a laser diode (LD, ID Photonics CoBriteDX1-1-C-H01-FA) centered at 1550.053 nm is injected into a dual-parallel Mach–Zehnder modulator (DP-MZM, Fujitsu FTM7960EX301) via a polarization controller (PC). The received signal in one receiving antenna, including the MIMO multipath SI signal and the SOI, is generated from an arbitrary wave generator (AWG, Keysight M8195A) and applied to one radio-frequency (RF) port of the upper DD-MZM of the DP-MZM. The other RF port of the DD-MZM is driven by the reference signal which is constructed by the LS algorithm with adaptive order in the digital domain via MATLAB-based DSP and then also generated from the AWG. One RF port of the lower DD-MZM is driven by a 20-dBm local oscillator (LO) signal generated from a microwave signal generator (MSG, Agilent 83630B). The two DD-MZMs are both biased at the minimum transmission point (MITP): One is used to implement the SIC and the other is used to generate the optical LO signal for frequency conversion. Good cancellation performance can be achieved when the amplitude and the time delay of the reference signal are adjusted reasonably. The optical signal from the DP-MZM is injected into a photodetector (PD), and the downconverted intermediate frequency (IF) signal from the PD is converted into a digital signal through an oscilloscope (OSC, R&S RTO2032) and further processed in MATLAB to construct the reference signal.

Then, the LS algorithm with adaptive order in the proposed system is further introduced. Channel estimation using the LS algorithm must know the originally transmitted signal $x(n)$ and the received signal $y(n)$. In the proposed system, $x(n)$ corresponds to IF signals $x_1(n)$ and $x_2(n)$ generated in the digital domain, and $y(n)$ corresponds to the IF received signal $y_{IF}(n)$ captured by the OSC. The LS estimation is performed using a data block of $N=40,000$ points in our experiment, which is determined by a combination of the oscilloscope sampling rate of 10 GSa/s and the signal length of 4 µs. For a more general case in which m different transmitting antennas are employed, the vector representations of the $y_{IF}(n)$ with $N$ observed samples are defined as

$$\mathbf{y}_{IF} = \mathbf{r}_{IF} + \mathbf{s}_{IF} + \mathbf{w}, \tag{1}$$

with

$$\mathbf{y}_{IF} = \begin{bmatrix} y(n) & y(n+1) & \cdots & y(n+N-1) \end{bmatrix}^T, \tag{2}$$

where $\mathbf{r}_{IF}$ is the MIMO multipath SI signal, $\mathbf{s}_{IF}$ is the SOI and $\mathbf{w}$ is the noise. The estimated SI signal is expressed as

$$\hat{\mathbf{r}}_{IF} = \mathbf{X}\hat{\mathbf{h}} = \begin{bmatrix} \mathbf{X}_1 & \mathbf{X}_2 & \cdots & \mathbf{X}_m \end{bmatrix} \begin{bmatrix} \hat{\mathbf{h}}_1 & \hat{\mathbf{h}}_2 & \cdots & \hat{\mathbf{h}}_m \end{bmatrix}^T. \tag{3}$$

Here $\mathbf{X}_j$ is the matrix of $x_j$ defined as follows

$$\mathbf{X}_j = \begin{bmatrix} x_j(n) & x_j(n-1) & \cdots & x_j(n-L+1) \\ x_j(n+1) & x_j(n) & \cdots & x_j(n-L+2) \\ \vdots & \vdots & \ddots & \vdots \\ x_j(n+N-1) & x_j(n+N-2) & \cdots & x_j(n+N-L) \end{bmatrix}, \tag{4}$$

where $j=1, 2, \ldots, m$ and $L$ is the order of the LS algorithm. The error vector is defined as $\mathbf{e}_{IF} = \mathbf{y}_{IF} - \hat{\mathbf{r}}_{IF}$, and then the LS estimation of $\hat{\mathbf{h}}$ can be obtained by minimizing the error vector,

$$\begin{aligned} \hat{\mathbf{h}} &= \arg\min \|\mathbf{e}_{IF}\|^2 \\ &= \arg\min \|\mathbf{y}_{IF} - \hat{\mathbf{r}}_{IF}\|^2 \\ &= (\mathbf{X}^H\mathbf{X})^{-1}\mathbf{X}^H\mathbf{y}_{IF}. \end{aligned} \tag{5}$$

To realize the LS algorithm with adaptive order, the order is introduced into the error vectors as a variable,

$$\mathbf{e}_{L(i)} = \mathbf{y}_{IF} - \hat{\mathbf{r}}_{L(i)} = \mathbf{y}_{IF} - \mathbf{X}_L\hat{\mathbf{h}}_L, \tag{6}$$

$$\mathbf{e}_{L(i)-\Delta} = \mathbf{y}_{IF} - \hat{\mathbf{r}}_{L(i)-\Delta} = \mathbf{y}_{IF} - \mathbf{X}_{L(i)-\Delta}\hat{\mathbf{h}}_{L-\Delta}. \tag{7}$$

Here, $i$ represent the iteration number, and $L$ and $L–\Delta$ represent the order of the LS algorithm. $e_{\Delta(i)}$ is defined as

$$\mathrm{e}_{\Delta(i)} = \sum_{k=1}^{N} \mathbf{e}_{L(i)}^{2}(k) - \sum_{k=1}^{N} \mathbf{e}_{L(i)-\Delta}^{2}(k). \tag{8}$$

Then, the order of the LS algorithm $L(i)$ is updated according to

$$l_f(i+1) = \left(l_f(i) - \alpha\right) - \gamma \mathrm{e}_{\Delta(i)}, \tag{9}$$

and

$$L(i+1) = \begin{cases} \lfloor l_f(i) \rfloor, & \text{if } |L(i) - l_f(i)| > \delta \\ L(i), & \text{otherwise} \end{cases}, \tag{10}$$

where $\alpha$, $\delta$, and $\Delta$ are all positive integers, $\lfloor\ \rfloor$ is the floor operator, and $\gamma$ is the step size for the order adaptation. An appropriate order of the LS algorithm can be obtained by setting the parameters of the algorithm reasonably and after a certain number of iterations. Then, the reference signal is reconstructed and the SI signal is suppressed by using the LS algorithm with an appropriate order.

## 3. Experiment and results

To verify the feasibility of the proposed system, an experiment is conducted. The three parameters $\alpha$, $\delta$, and $\Delta$ in the LS algorithm with adaptive order are set to 10, 1, and 40, respectively. $\mathrm{e}_{\Delta(i)}$ is very large when the order of the LS algorithm is much smaller than the value of the optimal order, so $\gamma$ is set to be very small from the beginning of the iteration to avoid situations where the order of the LS algorithm jumps to a large value and does not converge. However, at the end of the iteration, the order of the LS algorithm is close to the value of the optimal order and the $\mathrm{e}_{\Delta(i)}$ is small, so $\gamma$ should be set to a larger value to keep the order of the LS algorithm convergent. Thus, $\gamma$ is set to increase linearly with the number of iterations within the set range in the experiment.

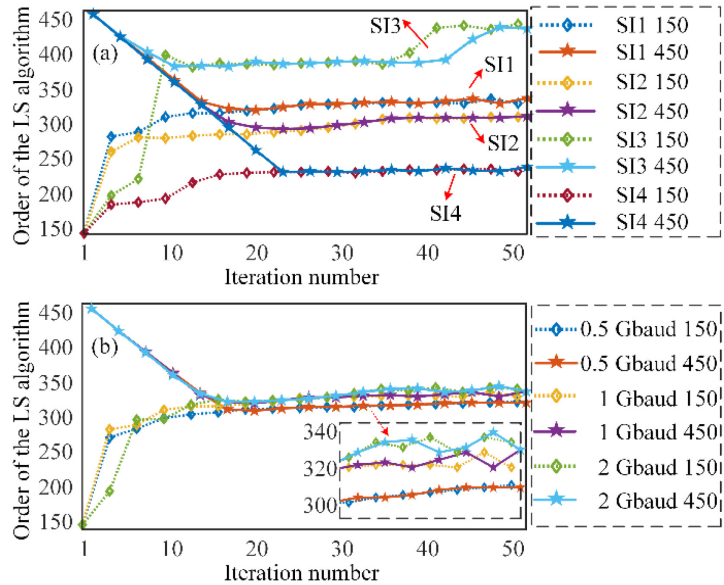

Fig. 2. The convergence of the LS algorithm with adaptive order in (a) different MIMO multipath cases, and in (b) different baud rates of SI1.

First, the convergence of the LS algorithm with adaptive order is investigated. Four different interference conditions are studied and the difference lies in the delay and amplitude of each SI path. In all the cases, the received signal contains the SOI and two different direct-path SI signals from two transmitting antennas and six multipath SI signals. The modulation format of all the signals used in the experiment is quadrature phase-shift keying (QPSK). The SI signals have a baud rate of 1 Gbaud and a carrier frequency of 10 GHz, while the SOI has the same carrier frequency and half the baud rate. In the experiment, the direct-path SI signal is fixed. The discrete path delay vectors of the SI signals related to the two direct-path SI signals in SI1 are [0 10 20 30] ns and [0 16 24 28] ns, and the corresponding discrete path gain vectors are [0 –3.09 –10.45 –20] dB and [0 –3.74 –9.12 –16.48] dB. The delay up to 30 ns corresponds to an order of 300 at a sampling rate of 10 GSa/s in the LS algorithm. As can be seen from Fig. 2(a), the order of the LS algorithm converges to near 320 whether iterated from 150 or 450. The order of the LS algorithm does not converge to 300 because of the initial parameter settings of the algorithm. Similar to the case of SI1, the maximums of the discrete delays in the cases of SI2, SI3, and SI4 are set to 28, 40, and 21 ns, respectively. As shown in Fig. 2(a), at a sampling rate of 10 GSa/s, the order of the LS algorithm converges to about 300, 420, and 230, respectively, which are very close to the theoretical value. Furthermore, the convergence of the proposed LS algorithm with adaptive order is also validated under different baud rates when the discrete path delay and path gain vectors of the received signals are set to be the same as SI1. When the baud rate is 0.5, 1, and 2 Gbaud, regardless of whether the initial order of the LS algorithm is set at 150 or 450, all the orders of the LS algorithm eventually converge to 300 to 340. To sum up, the order of the proposed LS algorithm with adaptive order can track the variation of the multipath SI channel adaptively and converge to a value around the optimal order.

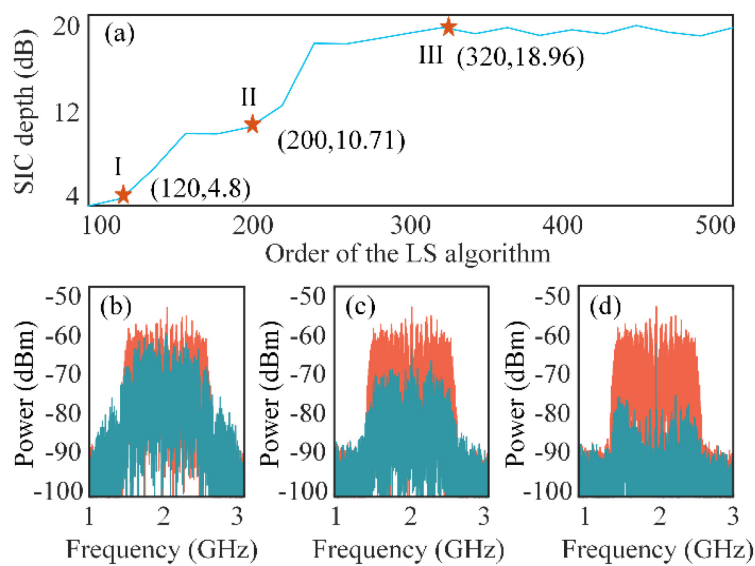

Fig. 3. (a) Variation of the SIC depth under different orders of LS algorithm when the MIMO multipath SI signal is in the case of SI1. Electrical spectra of the downconverted IF signal with and without SIC at point (b) I, (c) II, (d) III in (a).

Then, the influence of the order of the LS algorithm on the SIC depth is studied when the MIMO multipath SI signal is in the case of SI1 discussed above. To obtain the SIC depth accurately, the SOI is disabled in this study. The results are shown in Fig. 3(a). It can be seen that as the order of the LS algorithm increases from 100 to 300, the SIC depth keeps increasing. After the order exceeds the theoretical value of 300, the SIC depth remains basically the same and oscillates in a small range of around 19 dB. At point I, the order of the LS algorithm is 120, which is too small to include most multipath SI channel information in the estimation of the SI channel response. Thus, the reconstructed reference signal cannot suppress all the multipath SI signals and the SIC depth is only 4.8 dB as shown in Fig. 3(b). At point II, the order of the LS algorithm is 200. The estimation of the multipath SI channel response contains more information about the multipath SI channel and the SIC depth can reach 10.71 dB as shown in Fig. 3(c). At point III, the order of the LS algorithm is 320, and the estimation of the multipath SI channel response contains all the information of the multipath SI channel and the SIC depth can reach 18.96 dB as shown in Fig. 3(d). In theory, when the order of the LS algorithm is greater than the optimal value, the extra error in the estimation of the channel response obtained by using the LS algorithm increases with the increase of the order. However, the effect of the extra error on the SIC depth is very small compared to the effect of the error when the order of the LS algorithm is less than the optimal value, so the decrease of the SIC depth with the increase of the order over the optimal value is not seen in Fig. 3(a). It is worth noting that the discrete frequency component at 2 GHz shown in Fig. 3 is related to the AWG sampling rate and is introduced by the AWG. Furthermore, large fluctuations and fading can be observed in the spectra without SIC, which is due to the small power difference between the strongest multipath SI signal and the direct-path SI signal in the case of SI1.

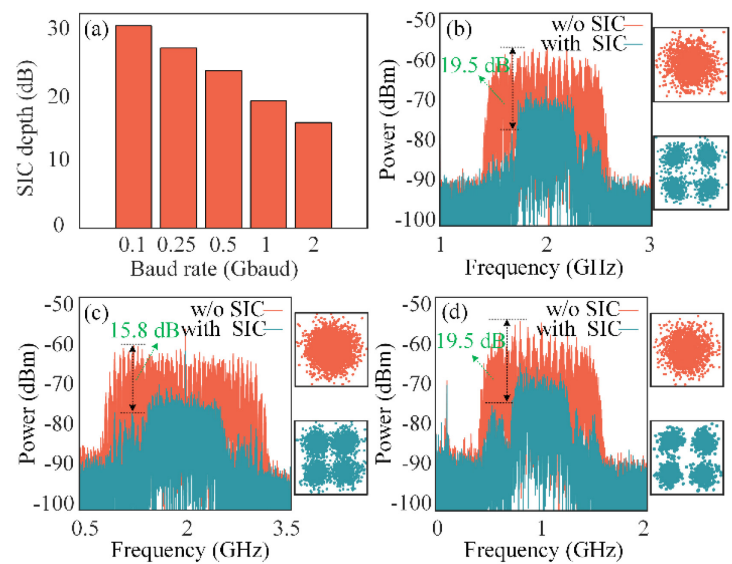

Fig. 4. (a) Variation of the SIC depth under different baud rates when the MIMO multipath SI signal is in the case of SI1. Electrical spectra and constellation diagrams of the downconverted IF signal with and without SIC when the carrier frequencies and baud rates are (b) 2 GHz, 1 Gbaud, (c) 2 GHz, 2 Gbaud, and (d) 1 GHz, 1 Gbaud.

Then, taking the SOI into account, the performance of the system is further studied. As shown in Fig. 4(a), when the discrete path delay and path gain vectors are set to be the same as SI1 discussed above and the frequency of the LO is 8 GHz, 30.2, 26.9, 23.5, 19.5, and 15.8 dB SIC depths are achieved when the SI signal has a carrier frequency of 10 GHz and baud rates of 0.1, 0.25, 0.5, 1, and 2 Gbaud, respectively. Figures 4(b) and 4(c) show the electrical spectra and the constellation diagrams of the downconverted IF signals in Fig. 4(a) with and without SIC when the baud rates are 1 and 2 Gbaud, respectively. When the SIC is turned off, the SOI is completely submerged by the spectrum of the SI signal, and the constellation diagrams of the SOI are chaos. However, when the reference signals are constructed by the LS algorithm with adaptive order and the SIC is enabled, 19.5, and 15.8 dB SIC depths are achieved. Furthermore, the constellations of the SOI become distinguishable from chaos and the EVMs of the recovered SOIs are 36.9% and 31.4% at the baud rates of 1 and 2 Gbaud, respectively. To verify the effectiveness of the system for the signals in different frequencies, the carrier frequency of the SI signal is set to 14 GHz, and the frequency of the LO is set to 13 GHz, so the frequency of the downconverted IF signal becomes 1 GHz. The corresponding spectra and constellation diagrams are shown in Fig. 4(d). An SIC depth of 19.5 dB is achieved and the EVM of the SOI is 29.8% after SIC, which is similar to the result when the carrier frequency is 10 GHz.

## 4. Conclusion

In summary, we have proposed and experimentally demonstrated a digitally-assisted photonic analog domain SIC and frequency downconversion method for the IBFD MIMO systems using the LS algorithm with adaptive order. The key significance of the work was the introduction of the LS algorithm with adaptive order to set the order of the LS algorithm reasonably, which balanced the performance and complexity of the algorithm used to reconstruct the reference signal. Moreover, photonics-assisted SIC was first demonstrated for IBFD MIMO systems with a more complex interference situation. Experiment results showed that the order of the LS algorithm proposed in this paper converged to appropriate values under various MIMO multipath SI channels, and 30.2, 26.9, 23.5, 19.5, and 15.8 dB cancellation depths were achieved when the SI signal had a carrier frequency of 10 GHz and baud rates of 0.1, 0.25, 0.5, 1, and 2 Gbaud, respectively.

**Acknowledgements**
This work was supported by the National Natural Science Foundation of China [grant number 61971193]; the Natural Science Foundation of Shanghai [grant number 20ZR1416100]; the Science and Technology Commission of Shanghai Municipality [grant number 18DZ2270800].

**References**
1. Z. Zhang, X. Chai, K. Long, A. V. Vasilakos, and L. Hanzo, "Full-duplex techniques for 5G networks: self-interference cancellation, protocol design, and relay selection," *IEEE Commun. Mag.*, vol. 53, no. 5, pp. 128–137, May 2015.

2. A. Sabharwal, P. Schniter, D. Guo, D. W. Bliss, S. Rangarajan, and R. Wichman, "In-band full-duplex wireless: challenges and opportunities," *IEEE J. Sel. Areas Commun.*, vol. 32, no. 9, pp. 1637–1652, Sep. 2014.

3. K. E. Kolodziej, B. T. Perry, and J. S. Herd, "In-band full-duplex technology: techniques and systems survey," *IEEE Trans. Microw. Theory Tech.*, vol. 67, no. 7, pp. 3025–3041, Jul. 2019.

4. A. T. Le, L. C. Tran, X. Huang, and Y. J. Guo, "Beam-based analog self-interference cancellation in full-duplex MIMO systems," *IEEE Trans. Wirel. Commun.*, vol. 19, no. 4, pp. 2460–2471, Apr. 2020.

5. E. Everett, A. Sahai, and A. Sabharwal, "Passive self-interference suppression for full-duplex infrastructure nodes," *IEEE Trans. Wirel. Commun.*, vol. 13, no. 2, pp. 680–694, Jan. 2014.

6. J. Yao, "Microwave photonics," *J. Lightw. Technol.*, vol. 27, no. 3, pp. 314–335, Feb. 2009.

7. Y. Chen, "A photonic-based wideband RF self-interference cancellation approach with fiber dispersion immunity," *J. Lightw. Technol.*, vol. 38, no. 17, pp. 4618–4624, Sep. 2020.

8. X. Han, B. Huo, Y. Shao, C. Wang, and M. Zhao, "RF self-interference cancellation using phase modulation and optical sideband filtering," *IEEE Photon. Technol. Lett.*, vol. 29, no. 11, pp. 917–920, Jun. 2017.

9. Y. Chen and S. Pan, "Simultaneous wideband radio-frequency self-interference cancellation and frequency downconversion for in-band full-duplex radio-over-fiber systems," *Opt. Lett.*, vol. 43, no. 13, pp. 3124–3127, Jul. 2018.

10. D. Zhu, X. Hu, W. Chen, D. Ben, and S. Pan, "Photonics-enabled simultaneous self-interference cancellation and image-reject mixing," *Opt. Lett.*, vol. 44, no. 22, pp. 5541–5544, Nov. 2019.

11. X. Fan, J. Du, G. Li, M. Li, N. Zhu, and W. Li, "RF self-interference cancellation and frequency downconversion with immunity to power fading based on optoelectronic oscillation," *J. Lightw. Technol.*, vol. 40, no. 12, pp. 3614–3621, Jun. 2022.

12. W. Zhou, P. Xiang, Z. Niu, M. Wang, and S. Pan, "Wideband optical multipath interference cancellation based on a dispersive element," *IEEE Photon. Technol. Lett.*, vol. 28, no. 8, pp. 849–851, Apr. 2016.

13. J. Chang, and P. R. Prucnal, "A novel analog photonic method for broadband multipath interference cancellation," *IEEE Microw. Wirel. Compon. Lett.*, vol. 23, no. 7, pp. 377–379, Jul. 2013.

14. X. Su, X. Han, S. Fu, S. Wang, C. Li, Q. Tan, G. Zhu, C. Wang, Z. Wu, Y. Gu, and M. Zhao, "Optical multipath RF self-interference cancellation based on phase modulation for full-duplex communication," *IEEE Photonics J.*, vol. 12, no. 4, Aug. 2020, Art. no. 7102114.

15. Y. Zhang, L. Li, S. Xiao, M. Bi, L. Huang, L. Zheng, and W. Hu, "EML-based multi-path self-interference cancellation with adaptive frequency-domain pre-equalization," *IEEE Photon. Technol. Lett.*, vol. 30, no. 12, pp. 1103–1106, Jun. 2018.

16. L. Zheng, Z. Liu, S. Xiao, M. P. Fok, Z. Zhang, and W. Hu, "Hybrid wideband multipath self-interference cancellation with an LMS pre-adaptive filter for in-band full-duplex OFDM signal transmission," *Opt. Lett.*, vol. 45, no. 23, pp. 6382–6385, Dec. 2020.

17. M. Han, T. Shi, and Y. Chen, "Digital-assisted photonic analog wideband multipath self-interference cancellation", *IEEE Photon. Technol. Lett.*, vol. 34, no. 5, pp. 299–302, Mar. 2022.